\documentclass[fleqn,usenatbib]{flesch}

\usepackage{mathptmx}
\usepackage[T1]{fontenc}

\usepackage{graphicx}	
\usepackage{amsmath}	
\usepackage{amssymb}	

\title[Milliquas v7.2]{The Million Quasars (Milliquas) v7.2 Catalogue, now with VLASS associations.  The inclusion of SDSS-DR16Q quasars is detailed.}

\author[E. Flesch]{\textbf{Eric Wim Flesch }$^{1}$\thanks{E-mail: eric@flesch.org}
\\
$^{1}$PO Box 15, Dannevirke 4942, New Zealand}

\pubyear{2021}

\begin{document}
\label{firstpage}
\pagerange{\pageref{firstpage}--\pageref{lastpage}}
\maketitle

\begin{abstract}
Announcing the release v7.2 of the Milliquas (Million Quasars) catalogue which presents all published quasars to 30 April 2021, including VLASS radio associations for the first time, and concluding the audit of quasars from SDSS-DR16Q and earlier SDSS releases.  The totals are 829\,666 classified type-I QSOs/AGN, 703\,348 quasar candidates of 60\%-100\% pQSO, plus type-II objects and blazars which bring the total count to 1\,573\,824.  Radio and/or X-ray associations, including probable double radio lobes, are shown for 333\,638 entries.  \textit{Gaia}-DR2 astrometry is given for most objects, as available.  The catalogue is available on multiple sites.

The inclusion of the SDSS-DR16Q quasars was a complex task with emergent issues which resulted in 13\,443 DR16Q entries being dropped, 1.79\% of their total.  There are also 1701 quasars included from earlier visual SDSS releases, as well as 14\,232 quasars from the SDSS-DR16 pipeline catalogue.  All these are explained here, including the validation of 677 additional high-redshift (z$\geq$3.5) SDSS quasars which were not included in DR16Q.    

\end{abstract}



\section{Introduction}
Milliquas v7.2 presents all quasars to 30 April 2021 from published papers large \& small.  The formulation of the Milliquas content and format is as given in the Half Million Quasars catalog \citep[HMQ:][]{HMQ} and references therein, but there's been some evolution since then.  Milliquas v6.4 \citep{FL2019} documented changes to the optical background data and numerous issues and data tweaks since the published HMQ.  This paper, in Section 2, summarizes subsequent changes and fixes since v6.4 and prior to this version, v7.2. 

\citet{FLES40} presented star spikes and asteroids found within the SDSS-DR16Q data, and more of those are given here, with images.  They serve as an introduction to the wider topic of identifying low-purity subsets within DR16Q, amounting to 1.79\% of their total, which are not accepted into Milliquas.  This procedure is described and enumerated in full in Section 3.  

SDSS-classified quasars which were not included in DR16Q are also considered, these being from earlier visual catalogues and the DR16 pipeline catalogue, with 14\,232 accepted into Milliquas.  Of special interest among those are 677 high-redshift (z$\geq$3.5) quasars with convincing spectra; these are detailed in Sections 4.2.2 and 4.5.1. 

This version is the first to show VLASS associations, these being calculated from their recent "Quick Look" catalogue; a quick round-up is given in Section 7.  The Milliquas v7.2 footprint on the sky is shown on Fig. 1; the SDSS footprints dominate, of course.

\begin{figure} 
\includegraphics[scale=0.3, angle=0]{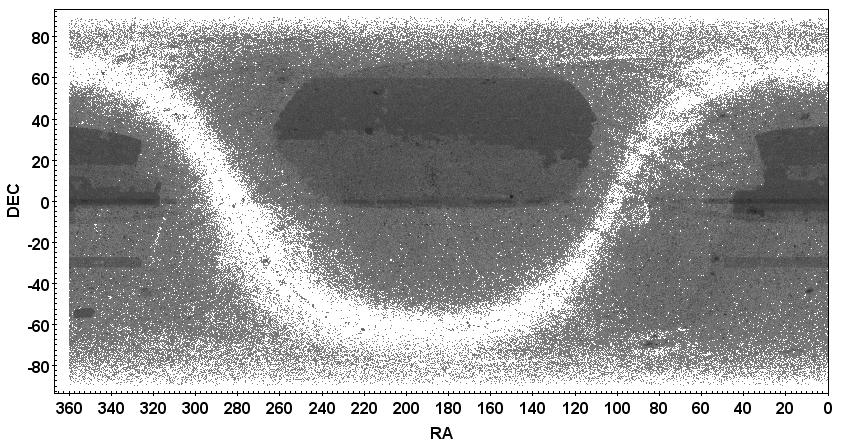} 
\caption{Sky coverage of Milliquas v7.2, includes classified objects and candidates.}
\end{figure}

Milliquas v7.2 can be downloaded from NASA HEASARC\footnote{https://heasarc.gsfc.nasa.gov/W3Browse/all/milliquas.html} or CDS VizieR\footnote{https://vizier.u-strasbg.fr/viz-bin/VizieR-3?-source=VII/290} which provide query pages, or from the Milliquas home page\footnote{http://quasars.org/milliquas.htm}.  Its ReadMe gives essential information about the data, including full citations.

\section{Changes and fixes in the past 2 years.} 

SDSS-DR16Q represented the conclusion of the eBOSS and BOSS quasar surveys, and that conclusion largely orphaned the leftover candidates from the SDSS-based photometric catalogues NBCKDE, NBCKDEv3, XDQSO, and Peters (citations given in the ReadMe).  My investigation showed significant anti-selection in those candidates, as many had been rejected by SDSS at the time of deciding BOSS/eBOSS targets.  Many others are too optically faint to be investigated in the medium term.  Thus for efficiency I dropped all said candidates which had no radio/X-ray/WISEA association, 574\,538 of them.  Said candidates which did have WISEA associations, numbering 35\,086, were renamed to display AllWISE names.

For quasar candidates, the pQSO (likelihood that the candidate is a quasar) threshold for inclusion into Milliquas has been lowered to 60\%. This is to present more radio/X-ray associated candidates which are not available elsewhere.  However, the pQSO$<$80\% candidates are not as reliable; close neighbours or unseen background sources often intrude. 

Pan-STARRS-based photometric redshifts were calculated for radio/X-ray associated candidates on the PS1 footprint (all-sky $\delta\geq-30^{\circ}$) using the four-colour method of \citet{HMQ}, Appendix 2.  About 1500 objects had their photometry supplemented from Pan-STARRS data.

LAMOST-DR6 was added.  Its new quasars are pipeline-only data which have not yet been evaluated visually by its authors, as earlier editions were.  Accordingly, its quasar classifications are accepted into Milliquas only when supported by radio/X-ray/WISEA pQSOs.  

The 4XMM-DR10 XMM-Newton catalog was added and X-ray associations calculated, replacing earlier editions.  The 2SXPS Swift XRT catalog was added and X-ray associations calculated, replacing 1SXPS.  

Gaia-DR2 \citep{Gaia} parallaxes and proper motions were analyzed to see if they could identify false quasars, but a 2\% false flag rate (usually from confusing close objects) showed that individual objects could not be disqualified in this way.  Legacy quasar publications were then group tested such that those with a high flag rate ($>$20\%), signaling the presence of low-quality data, had their flagged objects dropped, provided, as a confirmation, that those objects also had no radio/X-ray associations. The counts of dropped objects were:
\begin{itemize}
    \item  28 out of 40  (70\%) from Zhan \& Chen, 1987 \& 1989 ChA\&A
    \item 400 out of 916 (44\%) from Iovino/Clowes/Shaver, 1996,A\&AS,119,265
    \item  50 out of 194 (26\%) from Savage/Trew/Chen/Weston, 1984,MNRAS,207,393
    \item 111 out of 517 (21\%) from Drinkwater 1987.
\end{itemize}
Thus in total, 589 objects were dropped in this way.

\section{Reduction of the SDSS-DR16 Quasar catalogue} 

The Sloan Digital Sky Survey (SDSS) Quasar Catalogue 16th Data Release \citep[DR16Q:][]{DR16Q} features a main catalogue (DR16Q\_v4.csv) of 750\,414 quasars which includes earlier SDSS-I/II/III data.  About half of the DR16Q objects' spectra were visually evaluated, the remainder take classifications \& redshifts from the DR16 pipeline \citep[DR16:][]{DR16}.  The DR16Q processing emphasizes data uniformity and rule-based classifications so as to provide well-defined data for users.  Still, such a technique of bulk classification, half of it without visual inspection, presents a major challenge to the Milliquas scope of individual vetting of input quasars.  

The DR16Q paper abstract states an expectation of ``0.3\%-1.3\% contamination'' in the main catalogue, thus $\approx$99\% purity which matches my own threshold of inclusion into Milliquas.  So in principle I could simply incorporate the whole DR16Q into Milliquas.  But that would be a thoughtless method not consistent with the individual handling heretofore done in Milliquas.  More specifically, in a recent paper \citep{FLES40} I identified 82 DR16Q entries as non-quasars; naturally, I won't include those into Milliquas.  With that to break the ice, a larger consideration is to evaluate low-SNR pipeline-classified spectra which have various ZWARNING\footnote{ZWARNING is a DR16 binary field which flags potential problems with the pipeline spectroscopic processing.} values, especially ZWARNING=4 (more precisely, ZWARNING AND 4, ``small delta chi'') which flags when the classification and/or redshift was narrowly decided, even (in extreme cases) to arbitrarily small $\Delta\chi$ separating multiple contending hypotheses \citep[page 6]{Bolton}.  In plainer terms, well-understood spectral groupings, which yield reliable classifications and redshifts, are separated from eachother by transitional zones in which resident objects are of unclear nature -- and so flagged with ZWARNING=4.  An alternative approach would have been to not include such objects due to unclear classification or redshift, but DR16Q chose to include them for completeness, with the ZWARNING=4 flag.

By contrast, the Milliquas priority is purity alone, i.e., that classified QSOs should be 99\%+ reliably true QSOs.  It is a different scope from that of DR16Q, and so I need to identify subsets of low purity in DR16Q and exclude those from Milliquas.  At the start, the expectation was that there would be no such easily-identifiable low-purity subsets.  However, this notion was refuted by the following explicit subset:      

\begin{itemize}
\item The DR16Q data has a field IS\_QSO\_FINAL which flags accepted-quasar status; of those, 749749 rows have IS\_QSO\_FINAL=1 and 665 have IS\_QSO\_FINAL=2.  Of the latter, the DR16Q datamodel\footnote{commentary on IS\_QSO\_FINAL is at bottom of \scriptsize{https://data.sdss.org/datamodel/files/BOSS\%5FQSO/DR16Q/DR16Q\%5Fv4.html}} states it \textit{``means the object was questionably a quasar, but was kept. We recommend inspecting these objects before use.''}, which is a disclaimer much like "use at your own risk".  
\end{itemize}

I was surprised that such objects explicitly stated to be questionable and not passed fit by their authors would be included in DR16Q; also they are all flagged with Z\_CONF=1, i.e., low visual-inspection confidence.  Of course I was not going to try to confirm objects which the DR16Q authors themselves would not; accordingly, I dropped these 665 from the DR16Q intake, but these and other bulk-dropped data are still considered in the pipeline DR16 processing described in Section 4.5.  48 of these 665 do appear in Milliquas via either the pipeline processing or from legacy publications.      
   
With this as precedent, it is to be noted that about half of the DR16Q entries are flagged as visually inspected, and the remainder are taken from the DR16 pipeline.  I have applied rules documented in HMQ to subdivide the pipeline-sourced objects, in particular using the ZWARNING=4 flag and whether or not the pipeline SUBCLASS is populated.  As discussed in HMQ Section 3, the pipeline-provided SUBCLASS is taken as a key indicator of spectrum quality just on the basis of whether it is populated or not; an unpopulated SUBCLASS signals a spectrum of low significance because the pipeline could not decide subclass information.  This field isn't provided by DR16Q, but is easily obtained from the pipeline DR16 data.   DR16Q do not use it in their classification analysis.

By subdividing up the DR16Q data using such criteria different to that used by the DR16Q authors, I was able to identify subsets of low purity which I excluded from Milliquas.  The following text characterizes and ennumerates those exclusions; they are done sequentially in the order presented, so any dropped object is not further considered in the DR16Q intake.  Accordingly, the 665 DR16Q entries having IS\_QSO\_FINAL=2, discussed above, are not considered further in the following section.

\subsection{DR16Q entries seen to be false, or of low confidence or low significance}

\citet{FLES40} identified 82 false DR16Q entries as star spikes and other types, plus we have 2 swap-outs from the "New Quasars" section below (Fig. 11).  Also there are 34 additional false DR16Q entries not reported in \citet{FLES40}: these are 30 star spikes seen on Fig. 2 and 4 asteroids shown by Fig. 3, which were found amongst the low-confidence data discussed below, which were not considered for the objects of \citet{FLES40} which primarily dealt with medium-to-high confidence data.  All these total to 118 false entries which we permanently drop from all SDSS-DR16Q input data. 

358 DR16Q objects are flagged with CLASS\_PERSON=1, i.e., a visual inspector classified the object as a star, or as CLASS\_PERSON=4, i.e., visually classified as a galaxy or just ``not a quasar''.  These all show low-SNR spectra difficult of interpretation.  27 have Z\_CONF=3 (high-confidence visual classification) and 3 have Z\_CONF=2 (medium confidence), the rest are of low confidence.  Why they were included into DR16Q is unclear.  These 358 entries are dropped, but 31 of those do appear in Milliquas via either pipeline processing or legacy publications. 

The DR16Q data include, as a resource, QuasarNET \citep{QuasarNET} classifications and redshifts, for which its authors claim high (99\%) efficiency and accuracy.  Only 6944 ($\approx$1\%) of those QuasarNET classifications are as non-quasars.  Of those 6944 DR16Q objects, 4 are flagged with CLASS\_PERSON=50, i.e., blazar candidate without a redshift, and all 4 have ZWARNING=4.  4 others are pipeline-only objects of redshift<0.03.  These 8 marginal objects are of no interest and are dropped except for one blazar (J163651.46+262656.7) which is kept because it has radio/X-ray associations.  

\begin{figure} 
\includegraphics[scale=0.325, angle=0]{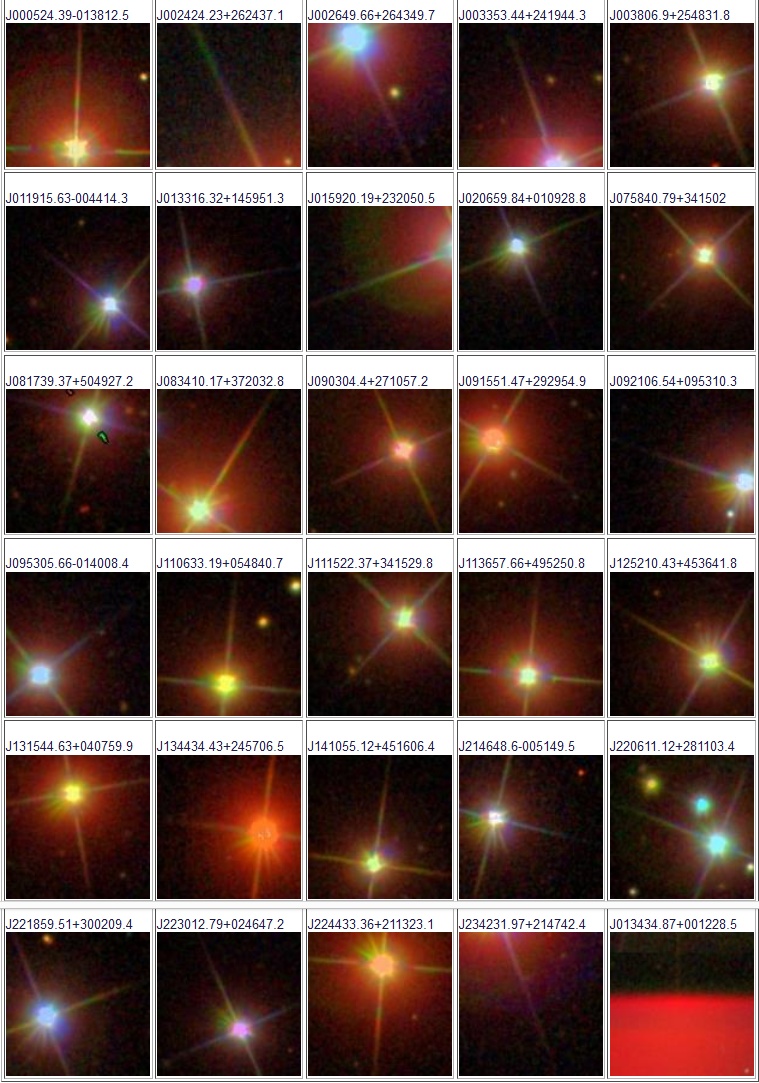} 
\caption{29 star spikes and 1 star glow artefact found amongst low-confidence DR16Q entries, images have 1 arcmin edges.  Annotated location is at exact centre of image, crossed by a star spike there with no background object present.}
\end{figure}
  
\begin{figure} 
\includegraphics[scale=0.325, angle=0]{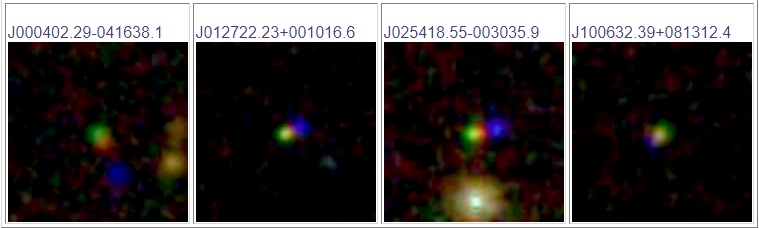} 
\caption{4 asteroids found amongst low-confidence DR16Q entries, images have 20 arcsec edges.  The different colours alternate as the moving object crosses the field of view during the successive \textit{ugriz} exposures.}
\end{figure}

3467 DR16Q entries not excluded above are of Z\_CONF=1 (low-confidence visual classification/redshift), and of those, 985 are in disagreement with QuasarNET as to quasar classification or redshift (median $\Delta$z=0.91, formally $\Delta$z/z$>$0.05).  Their spectra show no clear lines and 62\% have ZWARNING=4.  I find little evidence that any are quasars, and the 34 false objects of Figs. 2 \& 3 were amongst them.  Those 985 don't qualify to the Milliquas standard of 99\% quasar likelihood -- my estimate is well below 50\% -- and so are dropped.  However, 81 of those do appear in Milliquas via pipeline processing or legacy publications.  

About half of DR16Q objects source their redshift from the DR16 pipeline (SOURCE\_Z=PIPE) data which however can present multiple spectra for an object; in such cases the DR16Q selects the ``science primary''\footnote{https://www.sdss.org/dr16/spectro/catalogs/\#Selectinguniquespectra} spectrum.  However, DR16Q used an earlier version of the DR16 pipeline than the published one.  In 124 such cases the published DR16 pipeline presented a lower redshift than the version used by DR16Q, either because the spectrum was reprocessed or a different one chosen as the primary; 50 of those have $\Delta$z$>$2.  Those 124 are dropped from the input DR16Q processing so that their updated redshifts will be picked up in the input DR16 pipeline processing instead.  Of those, 71 are accepted into Milliquas; the high retention rate makes sense because the issue was just that of the timing of improved processing.        

Here's the big one:  those DR16Q entries which are pipeline-sourced (i.e., SOURCE\_Z=PIPE) and for which the pipeline-provided SUBCLASS is not populated (thus indicating a low-information spectrum), and furthermore have the ZWARNING=4 flag, i.e., their classifications/redshifts were narrowly decided, show unclear spectra of low significance which are difficult of interpretation.  They amount to 12\,228 objects for which my visual estimate is that less than half are true quasars -- there are no clear quasar lines seen in those spectra.  They are well below the Milliquas standard of 99\% quasar likelihood.  Therefore all such DR16Q entries are dropped (and so not further considered below), but 983 of those do appear in Milliquas via legacy publications or radio/X-ray/WISEA associations onto pipeline objects. 

Thus so far 14\,486 DR16Q entries have been dropped from this processing, but 1215 of them do appear in Milliquas from legacy publications or DR16 pipeline processing.  Thus the net total DR16Q entries dropped so far is 13\,271.  A few small drops will be added below.  

\subsection{High-redshift quasars}

DR16Q production treated z$\geq$3.5 ("hi-z") as a high-redshift category which required visual confirmation to be accepted.  The actual count is 10\,370 hi-z entries which, after the exclusions of Sec. 3.1, leave 9626 hi-z entries of which 9038 are visually inspected, 287 are inherited from the visually-inspected DR7Q/DR12Q, and 301 are from the DR16 pipeline.  High-redshift spectra are particularly recognizable by the prominent Ly$\alpha$ line with the declining flux profile blueward of it (the Lyman forest) which is bounded by the Lyman cutoff at $\lambda_{cutoff}=\frac{3}{4}\lambda_{Ly\alpha}$ exactly, and crested by the Ly$\beta$ line at $\lambda_{Ly\beta}=\frac{27}{32}\lambda_{Ly\alpha}$ exactly (see Fig. 4).  In practice, the dropout and Ly$\beta$ locations are seen to vary due to noise and forest absorption, but this Lyman profile is the definitive feature by which to identify the hi-z sources.

The 9038 visually inspected hi-z entries consist of 8942 having Z\_CONF=3 (high confidence) of which 8932 (99.89\%) are accepted as hi-z quasars into Milliquas, 94 have Z\_CONF=2 (medium confidence) of which just 10 are accepted, and 2 have no Z\_CONF but have good spectra and are accepted.  The 94 dropped high-z objects show featureless or single-line spectra without any recognizable Lyman feature; 12 have legacy identifications as blazars or low-z quasars, the remaining 82 are mostly flagged with ZWARNING=4 and one is a star spike.  Therefore the total number of DR16Q visually-inspected z$\geq$3.5 quasars accepted into Milliquas as hi-z quasars is 8944, 12 are accepted as low-z objects, and 82 are dropped. 

287 hi-z quasars are inherited from DR7Q and one from DR12Q, all of which were visually confirmed in their day, but one object is bogus: J185319.41+180715.1 is a faint star 4 arcsec from a mag-18 white dwarf star in a dense star field, which somehow led to a mistaken classification.  So 286 inherited z$\geq$3.5 quasars are accepted into Milliquas, and 1 is dropped. 

For the 301 pipeline-sourced hi-z entries, the accompanying QuasarNET redshift is much lower for every case but 8; my visual inspection of those spectra supports the QuasarNET redshift values as clearly better.  I present the QuasarNET redshifts for those objects in Milliquas, flagged with zcite=``DR16QN'' to show the provenance.  The 8 high-redshift objects for which the QuasarNET and pipeline redshifts agree, all show convincing spectra with a dominant Ly$\alpha$ line and the Lyman forest \& Ly$\beta$ line blueward of it. 

Therefore, of the 9626 z$\geq$3.5 DR16Q entries, 9238 are accepted into Milliquas as hi-z quasars, 305 are accepted as lower-redshift quasars or blazars, and 83 are dropped as bad objects.  There is one other DR16Q object, J013413.11+134952.3 with a pipeline-sourced redshift of 3.491, which I switched to the QuasarNET redshift of 3.508 after inspecting the spectrum.  Therefore a total of 9239 DR16Q quasars appear in Milliquas as hi-z quasars, but 159 of those have legacy citations, so 9080 DR16Q-cited z$\geq$3.5 quasars are displayed in Milliquas.  

\begin{figure} 
\includegraphics[scale=0.24, angle=0]{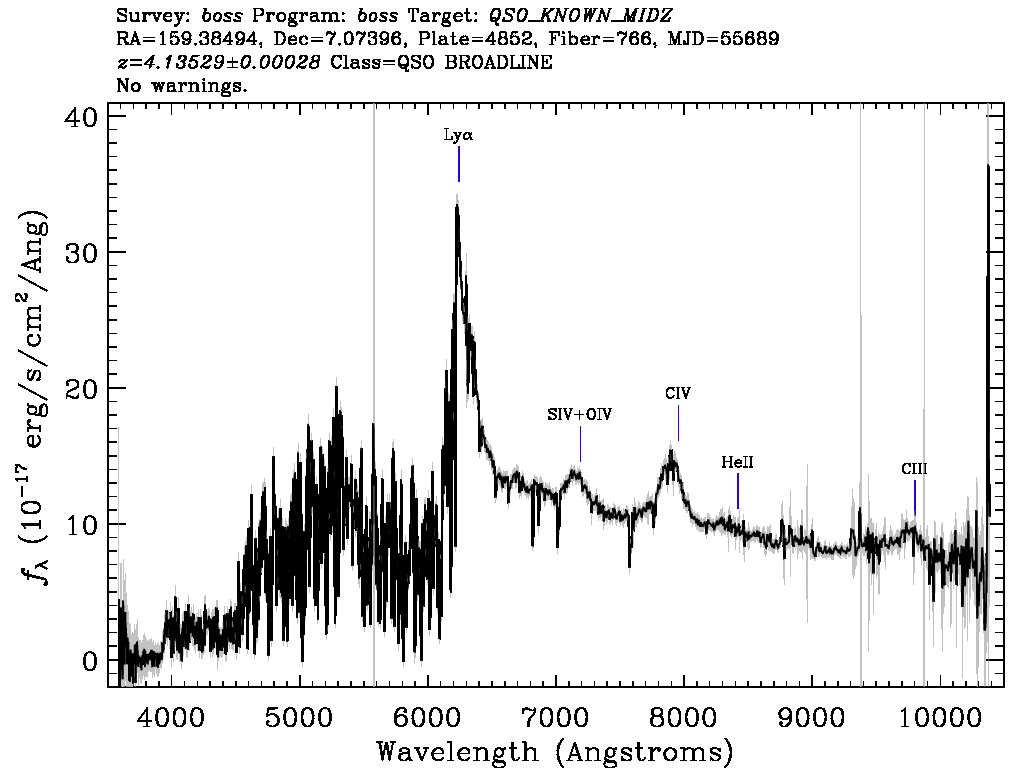} 
\caption{A high-redshift spectrum dominated by the Ly$\alpha$ line at 6250\AA\ at this redshift of z=4.135.  Blueward (leftward) of it is the Lyman forest with the Ly$\beta$ line at 5280\AA\ and the Lyman cutoff at about 4550\AA.  The dichroic seam at 5580\AA\ often shows a false line.}
\end{figure}

However, Section 4 gives 677 more z$>$3.5 SDSS quasars which are accepted into Milliquas but were not presented in DR16Q for unknown reasons.

\subsection{Wash-up}

This leaves 726\,301 DR16Q entries to consider, all of which are accepted into Milliquas barring 89 which are seen to have spectra of no merit.  

23 are faint (r$>$22.0) SEQUELS objects which are re-observed unclassified targets, those 23 evidently not helped by the 2nd attempt.  Their spectra are not flagged as visually inspected, which for SEQUELS objects (which are all visually inspected)  means just that they could not be visually classified, and show no recognizable feature, although assigned pipeline redshifts.  There are only 8 QuasarNET redshifts for these, and they disagree ($\Delta$z$>$0.5 mostly) with the pipeline redshifts in all cases.  They have no SUBCLASS but do have have ZWARNING=0 which is why they weren't dropped in Section 3.1.  

Nine are DR7Q-inherited objects which have since been pipeline-classified as stars and QuasarNET agreeing that they are not quasars.  Four have ZWARNING=4, one has proper motion, and their spectra are star-like or featureless.  They are J000219.33-105259.5, J081402.32+040508.9, J101449.56+032600.8, J122900.61+441958.6, J130201.30+092351.6, J152200.77+035016.9, J205707.67+752603.8, J212210.95-080611.7, and J212849.04+000447.6.

Similarly, 47 were visually inspected and classified as quasars with high or medium confidence, but pipeline-classified as stars and QuasarNET agreeing they are not quasars, except for 3 where QuasarNET gives different redshifts.  All but 5 have have ZWARNING=4, and their spectra are featureless /unrecognizable except for one obvious white dwarf (J142134.24+154204.4). 
       
Of the remaining 10, two (J002128.60+150223.7 and J110616.63+253941.5) were previously classified as white dwarfs and their spectra confirm that, one (J083208.97+472526.4 with pipeline z=2.369) is a line poacher\footnote{"line poachers" are explained in \citet{FLES40}, Section 5.} (onto J083209.11+472524.4, z=2.371), two were previously visually classified as stars, and the remainder show inscrutable spectra.   

All these total to 13443 net entries dropped from the DR16Q for a drop rate of 1.79\%, slightly in excess of its authors' estimate of 1.3\% contamination, but surely I have dropped many true quasars amongst the 12K inscrutable ZWARNING=4 entries -- if I dropped 4K true quasars, a plausible or probable scenario, then my count of false objects matches the DR16Q authors' estimate.  It is just that my priority is purity of the data whereas the DR16Q authors also favoured completeness.  However, in the next section I identify an additional 15\,531 SDSS quasars which were left out of the DR16Q but are included in Milliquas. 

In total, 736\,971 DR16Q quasars are in Milliquas, with 715\,815 so cited, and 21\,156 with other citations.  Also 12 DR16Q entries are presented as candidates in Milliquas, showing radio associations but with pQSO$<$97\%, and there are 71 DR16Q entries classified as type-II objects by \citet{YSZ}.   

It is to be mentioned that when the QuasarNET redshift value is much lower than the pipeline redshift presented by DR16Q, then I present the QuasarNET redshift instead; inspection shows the QuasarNET redshifts are reliably better for this subset.  This has remedied 1259 too-high pipeline redshifts presented by DR16Q, including 396 which had been presented by DR16Q as hi-z (z$\geq$3.5) quasars, 217 of which had been presented as z$>$5 quasars.  

Also, as a note on the effect of using SUBCLASS as an important indicator, when an object has more than one pipeline spectrum, the general rule is to utilize the one designated as ``sciencePrimary''; but infrequently it happens that the sciencePrimary spectrum has no SUBCLASS but an alternative spectrum does have the populated SUBCLASS.  In such cases (57 of them), I select that alternative spectrum as the one to use.  Usually it makes little difference, but an example of where it did is J023946.49-025152.6 for which DR16Q gives z=2.568 and Milliquas gives z=1.811 which is clearly the better value upon inspection of those spectra.

\section{Inclusion of other SDSS quasars into Milliquas} 

DR16Q presents an impressive round-up of quasars classified by SDSS to the end of SDSS-IV, but it was not entirely comprehensive.  Some quasars from earlier visual catalogues were omitted, often by design to avoid taking up bad objects, and also to present a uniform catalogue which could be used onwards by researchers of all stripes.  Milliquas's sole goal is to include all published quasars which are assessed at pQSO$\geq$99\%, so all quasars from earlier SDSS quasar catalogues and all SDSS pipeline catalogues are considered, and a total of 15\,531 additional SDSS quasars are included into Milliquas, as follows.

\subsection{\textbf{The DR16Q SuperSet}}    

The SDSS-DR16Q Superset is the other dataset presented by DR16Q, it gives classifications for 1\,440\,615 galaxies, stars, and quasars observed over SDSS III/IV plus DR7 quasars.  Of its objects which are not already in the main DR16Q dataset, 241\,324 have visually inspected spectra classified with high confidence (Z\_CONF=3).  Of those, there are 40 objects classified as quasars, which are however anti-selected due to their not having been accepted into the main quasar catalog.  Inspection of their spectra shows that 19 have convincing quasar lines so they are included in Milliquas; these are cited with pipeline or legacy citations to avoid confusion with the main DR16Q dataset, and the redshift citation is to "DR16QN", i.e., the QuasarNET redshift.  The remaining 21 show non-quasar spectra and were classified as stars by the DR12Q superset.

\subsection{The DR14Q visual catalogue} 

The SDSS-SR14Q \citep{DR14Q} was the immediate predecessor of, and superseded by, DR16Q.  However, 1489 DR14Q objects which were not taken up by DR16Q are found to be valid and included in Milliquas.  These are analyzed here in groups.

\begin{figure} 
\includegraphics[scale=0.35, angle=0]{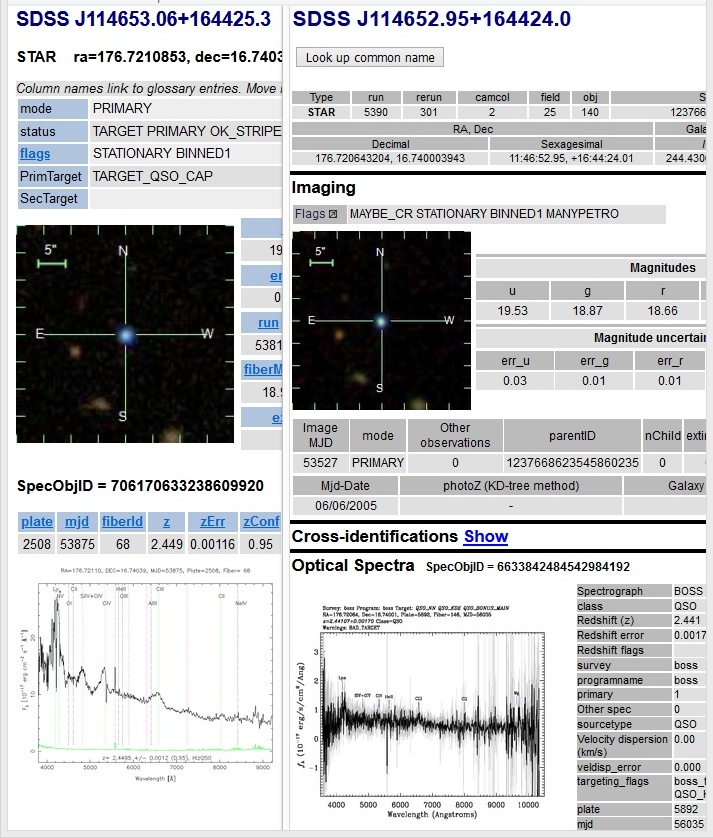} 
\caption{Bad Target: At left, the legacy DR7 finding chart of J114653.06+164425.3 which shows a spectrum of broad quasar lines dominated by the Ly$\alpha$ line at 4300\AA, establishing a redshift of 2.449.  At right, a subsequent \textit{BOSS} spectrum in a swath of sky with tracking slippage of 2.049 arcsec; the image is onto the slipped astrometry.  The spectrum shows residual bleed-over light from the missed quasar; just enough was captured to still enable a redshift measurement of 2.441.}
\end{figure}

\subsubsection{\textbf{So-Called BAD TARGETS}}

DR16Q elected not to include any objects flagged with ZWARNING=256 (precisely, bit 8 set) which means ``bad target'' or bad astrometry.  This was due to tracking problems in certain places during observing runs, so these objects come in astrometric swathes -- a total of 6228 sources so flagged in about 80 swathes, in the DR16 pipeline catalogue.  Among those are 1228 pipeline quasars of which 767 were presented as approved quasars by DR14Q.  Fig. 5 shows what bad tracking did to the astrometric accuracy of one quasar (not a member of the 767), and the resultant spectral degradation; the caption explains fully.  The astrometric offset between the 2 manifestations is 2.049 arcsec, beyond the standard 2-arcsec de-duplication, so DR14Q presented both of these as quasars.  DR16Q includes the left source only, dropping the right source due to its ``bad-target'' flag.    

\begin{figure} 
\includegraphics[scale=0.30, angle=0]{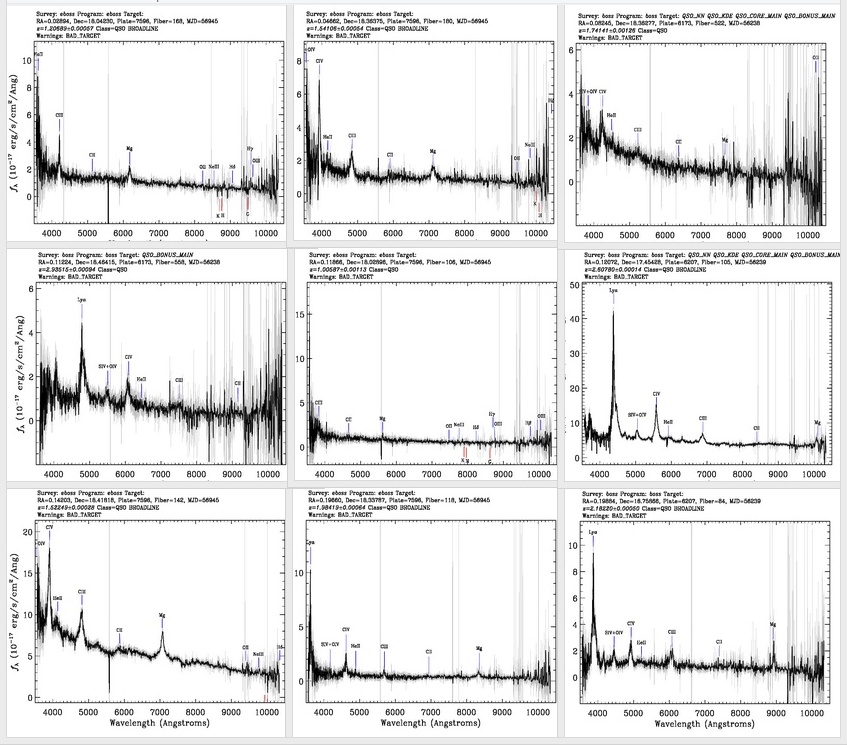} 
\caption{The first 9 of the 767 "bad target" DR14Q quasars, in RA order.  The centre spectrum shows the difficult redshift z=1.005.}
\end{figure}   

Perhaps DR16Q excluded all bad-target objects to prevent such beyond-2-arcsec duplications, and because off-target pointing gives skewed photometry.  The problem with that is that Fig. 5 shows the most extreme case only; all other DR14Q-presented ``bad target'' quasars were better, usually much better, pointed.  Their astrometry does not vary much from the true.  488 of those 767 match to Gaia-DR2 \citep{Gaia} (the remainder all have \textit{r}$>$20.1, 98\% \textit{r}$>$20.5, thus too faint for Gaia) and of those, 90\% match within 0.2 arcsec, 97\% within 0.5 arcsec, and 99\% within 1.0 arcsec; the farthest is 1.9 arcsec.  It looks like the baby got tossed out with the bathwater in a big way -- the 767 DR14Q bad-target quasars are all good valid objects and are included in Milliquas as the valid quasars that they are.  Fig. 6 displays the first 9 spectra of these 767 (in RA order) to show that these are reliable quasars.

\subsubsection{\textbf{High-Redshift Quasars}}    
       
As stated in Sec. 3.1, DR16Q treated z$\geq$3.5 ("hi-z") as a high-redshift category which required visual confirmation to be accepted.  They seemingly did not extend their inspections to the 410 DR14Q-only hi-z quasars, i.e., those not inherited from DR12Q, because 394 of those show valid hi-z spectra with almost all having ZWARNING=0 (i.e., no problems with the spectra), but were not included in DR16Q.  My visual inspection reveals those 394 have strong spectra in good accordance to the hi-z profile of Fig. 4, see Fig. 7 for the first 9 spectra of them (in RA order).  It seems an oversight that DR16Q did not visually inspect these important objects.  They are all in Milliquas, and can be retrieved by querying on cite="DR14Q" and z$\geq{3.5}$.

\begin{figure} 
\includegraphics[scale=0.30, angle=0]{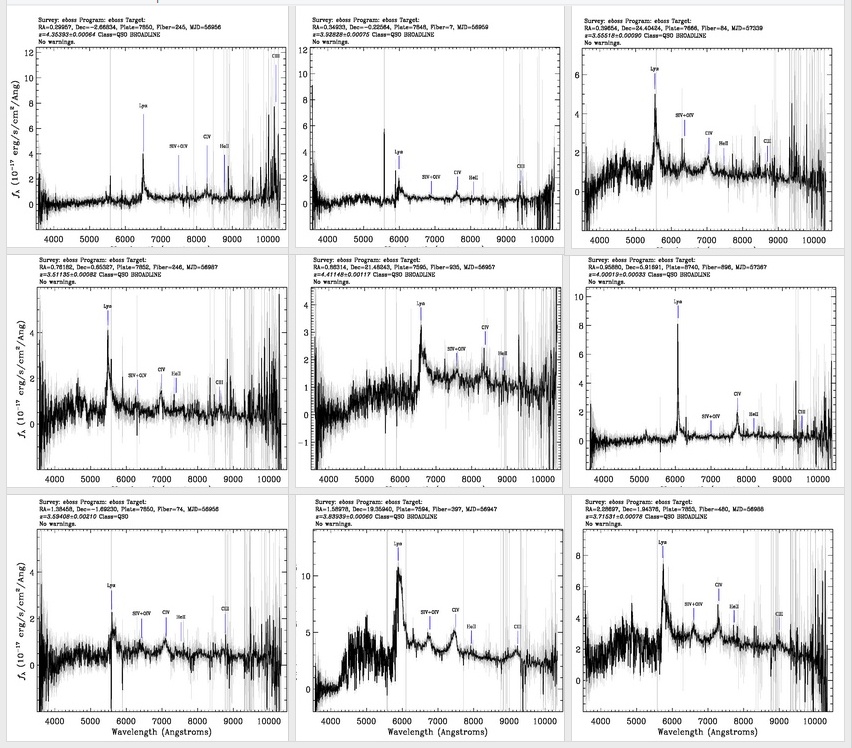} 
\caption{The first 9 of the 394 high-redshift DR14Q quasars, in RA order.  Note the prominent Ly$\alpha$ line and bluewards (left) of it, all or some of the Ly$\beta$ line, the Lyman cutoff, and the Lyman forest.}
\end{figure}             

15 more DR14Q hi-z quasars were inherited from DR12Q and are also absent from DR16Q in spite of good spectra.  Eight of them are also flagged as bad-target, their spectra are shown in Fig. 8; comparison to the hi-z profile of Fig. 4 shows these are all valid hi-z spectra.      

\begin{figure} 
\includegraphics[scale=0.22, angle=0]{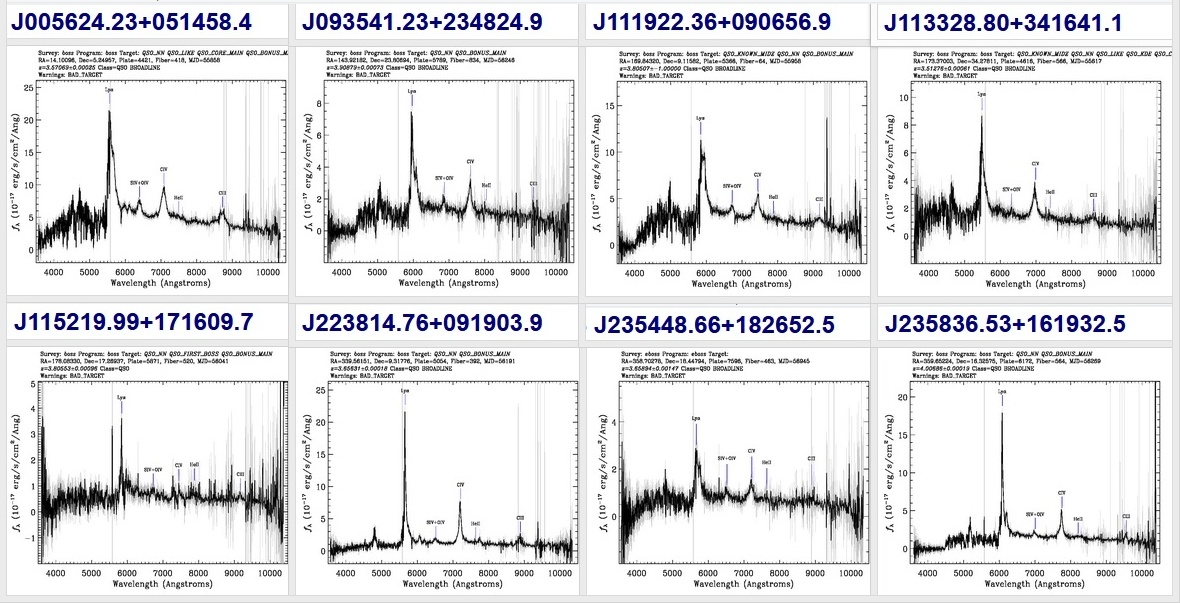} 
\caption{All 8 DR14Q spectra which are both z$>$3.5 and flagged as "BAD\_TARGET".  All spectra show the strong Ly$\alpha$ line and a well-placed Ly$\beta$ line at $\lambda_{Ly\beta}=\frac{27}{32}\lambda_{Ly\alpha}$ -- except for the 5$^{th}$ spectrum (at lower left) which suffers from spectral degradation due to an 0.8 arcsec offset.  That object has a radio detection, FIRST J115219.9+171610.  These quasars are not in DR16Q; they are included in Milliquas.  Zoom in to see clearly.}
\end{figure} 

4 of the DR14Q hi-z quasars inherited from DR12Q were flagged with ZWARNING=0, i.e., nothing wrong with them.  I've checked their spectra and they are fine.  It is puzzling that DR16Q missed these objects given that they prioritized the uptake from DR7Q and DR12Q.  The four quasars are included in Milliquas, they are:
\begin{itemize}
\item J092830.95+493141.2 with z=3.807  
\item J093558.36+225407.4 with z=3.751 
\item J112907.58+600240.6 with z=4.223 
\item J133544.95+263313.1 with z=3.706 
\end{itemize}

\subsubsection{\textbf{Miscellaneous notes}}  

19 DR14Q redshifts appear to come from re-observations because those spectra are not in the pipeline file.  The highest-redshift one is J093509.47+475255.9 with z=4.014 (befitting its photometry), the next highest is J131308.18+045945.6 with z=2.519, then J143619.34+000512.9 with z=2.372, and so on.  The DR14Q paper did not document the origins of these redshifts.  I have retained them trusting that they were special re-observations.

16 DR14Q quasars have ZWARNING=128 which is the UNPLUGGED flag which indicates that the CCDs were not firmly plugged or that the fibre had failed, or even that fibres had crossed to the wrong plugholes.  But if the spectrum matches the photometry both in flux and colour profile, then the spectrum will be correct albeit possibly slightly degraded.  My inspection shows that the spectra and photometry of all 16 objects are well-matched both in flux and colour profile, and they all show good quasar lines, so they are kept.  

583 of the DR14Q-only objects have ZWARNING=0, i.e., all OK, and they show quasar spectra of varying quality but all with quasar lines, so I keep those.  Also, there are 38 DR14Q quasars which DR16Q did report but with a higher redshift; all 38 bear QuasarNET redshifts which agree with the DR14Q value, so those are used and cited to DR14Q in Milliquas.

\subsection{The DR12Q visual catalogue} 

The SDSS-DR12Q \citep{DR12Q} is fully taken up by DR16Q except for some which did appear in DR14Q which are treated in the preceding section.   Also 235 DR12Q entries were classified as Type-II and 8 entries as blazars by other authors.  So there are no DR12Q-only quasars in Milliquas.

\subsection{The DR7Q visual catalogue} 

The SDSS-DR7Q \citep{DR7Q} was, in principle, fully taken up by DR16Q.  However, 1967 DR7Q quasars are presented by Milliquas.  Most of those (1774) are only because DR7Q presented legacy names for their quasars, names not given by the VCV \citep{VCV} quasar catalogue which was the predecessor of Milliquas.  Such legacy names are a priority, so Milliquas presents those names and the DR7Q citation for them.  However, this leaves 193 DR7Q quasars which were not taken up by DR16Q, in spite of their stated intention to do so.  

189 of the 193 come from the Table 5 of DR7Q which is presented in the Appendix A of that paper, with a total of 207 entries, 4 of which were duplicates, thus 203 unique entries.  Of those, 6 are in DR16Q so presumably re-surveyed by BOSS, and 5 are since found to be galaxies, 1 type-II galaxy, 1 white dwarf star (J084648.03+451258.7), and 1 object which appears not to exist (J024932.12-080814.9).  This leaves 189 \textit{bona fide} DR7Q quasars from its Table 5 which are not in DR16Q.  Perhaps the DR16Q authors didn't know about Table 5; it is no longer carried on the SDSS website.  However, it is still available on the publisher's page, and those quasars are in Milliquas.

This leaves 4 DR7Q-only quasars still unaccounted for.  They are:
\begin{itemize}
\item LBQS 0052-0015 with z=0.647 and X-ray CXO J005441.1+000109 and radio FIRST J005441.1+000110 
\item SDSS J014942.51+001501.7 with z=0.552 and X-ray 4XMM J014942.5+001501
\item SDSS J082012.63+431358.4 with z=1.073 and X-ray 2SXPS J082012.5+431357 and radio FIRST J082012.6+431358 
\item FBQS J113324.7+323449 with z=2.537 and radio FIRST J113324.7+323449 
\end{itemize}

These all have good QSO lines, with starforming continua present as well.  The first 3 dropped out of the visual catalogues after DR7Q, never to re-appear.  The last one did appear in DR12Q and DR14Q, but dropped out of DR16Q.  They have excellent radio/X-ray, and are in Milliquas.

\subsection{The DR16 pipeline catalogue} 

\begin{figure} 
\includegraphics[scale=0.30, angle=0]{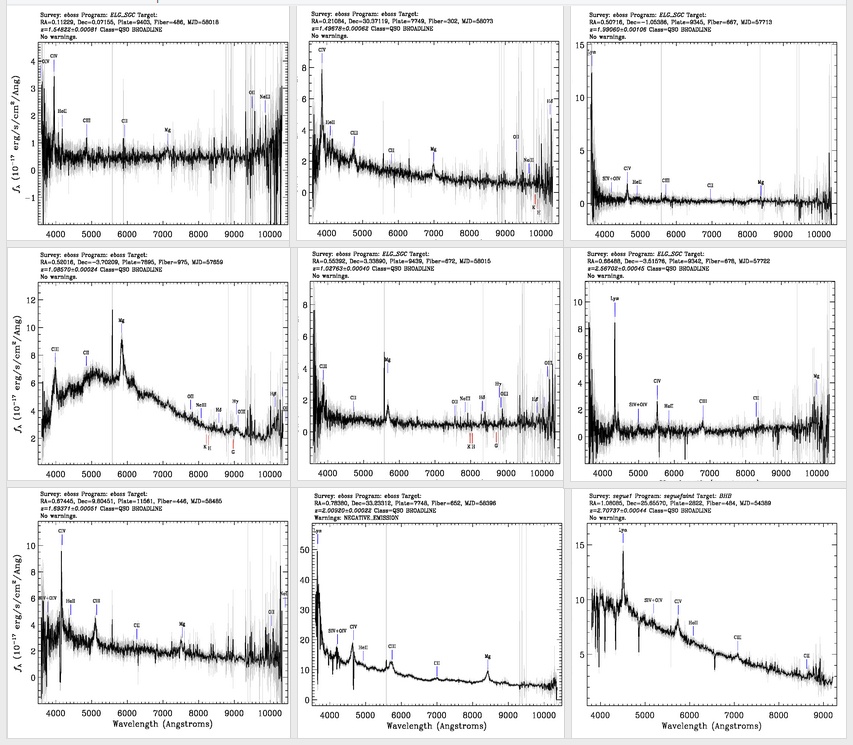} 
\caption{The first 9 of the 2330 z$\geq{1.0}$ DR16 pipeline quasars, in RA order.}
\end{figure}  

The SDSS-DR16 pipeline catalogue \citep[DR16:][]{DR16} gives pipeline-processed details of 5\,789\,200 spectra taken on the Sloan 2.5m Telescope over the life of the SDSS project, including multiple spectra for many targets.  It supersedes the previous SDSS pipeline catalogues and gives pipeline classificiations of quasars, galaxies, and stars via the latest algorithms.  These come to 4\,665\,607 unique targets which classify as 901\,824 quasars, 2\,784\,895 galaxies, and 978\,888 stars.  Most of the quasars are taken up by the DR16Q catalogue; of the remainder, I select those which have populated SUBCLASS and do not have ZWARNING=4, and also those with radio/X-ray/AllWISE associations.  This selection ensures we have only the highest-quality objects.  These amount to 31\,639 pipeline quasars which are, in principle, anti-selected by virtue of not having been included into DR16Q.  However, these objects are still useful, as follows. 

Of those 31\,639, 17\,407 are subclassed as \underline{not} BROADLINE, so I classify those as type-II quasars, although it doesn't exactly correspond to those.  True type-IIs have narrow beaming OIII lines or OII lines \footnote{One physical model for this is that the central singularity is orbited by a stratified torus of plasma, the outer shell of which dissipates the heat by radiating those lines.}, so to interpret not-BROADLINE as being type-II is an approximation at best.

13\,830 have the BROADLINE subclass (and 402 have no subclass) so are presumptive type-I quasars, but inspection of those with z$<$1 show that many, even most, are type-II objects because the beaming OIII/OII lines are present.  I haven't yet had the heart to individually inspect all those spectra, or find other criteria to subset them, so users are cautioned that the 11\,902 z$<$1 DR16 pipeline quasars are a mix of type-I and type-II objects.

This leaves 2330 DR16 pipeline quasars which have z$\geq{1}$, and these look reliable; Fig. 9 shows the first 9 of these in RA order, and clear quasar lines are seen throughout, even if they aren't the best-looking spectra.  I've checked a few hundred of them and all look fine, so they can be used with confidence.

\subsubsection{\textbf{High-Redshift Quasars}}

A subset of the z$\geq{1}$ pipeline quasars are the z$\geq$3.5 ("hi-z") quasars, a desirable subset as these are important objects for cosmology if valid; there are 284 hi-z objects amongst the pipeline entries.  I have visually inspected each spectrum, and 268 are found to be OK hi-z quasars and 16 were bogus and dropped.  The valid 268 have strong spectra in good accordance to the hi-z profile of Fig. 4, see Fig. 10 for the first 12 spectra of them (in RA order).  Again it is surprising that DR16Q did not visually inspect these important objects.  They are all in Milliquas, and can be retrieved by querying on cite="DR16" and z$\geq{3.5}$.

\begin{figure} 
\includegraphics[scale=0.30, angle=0]{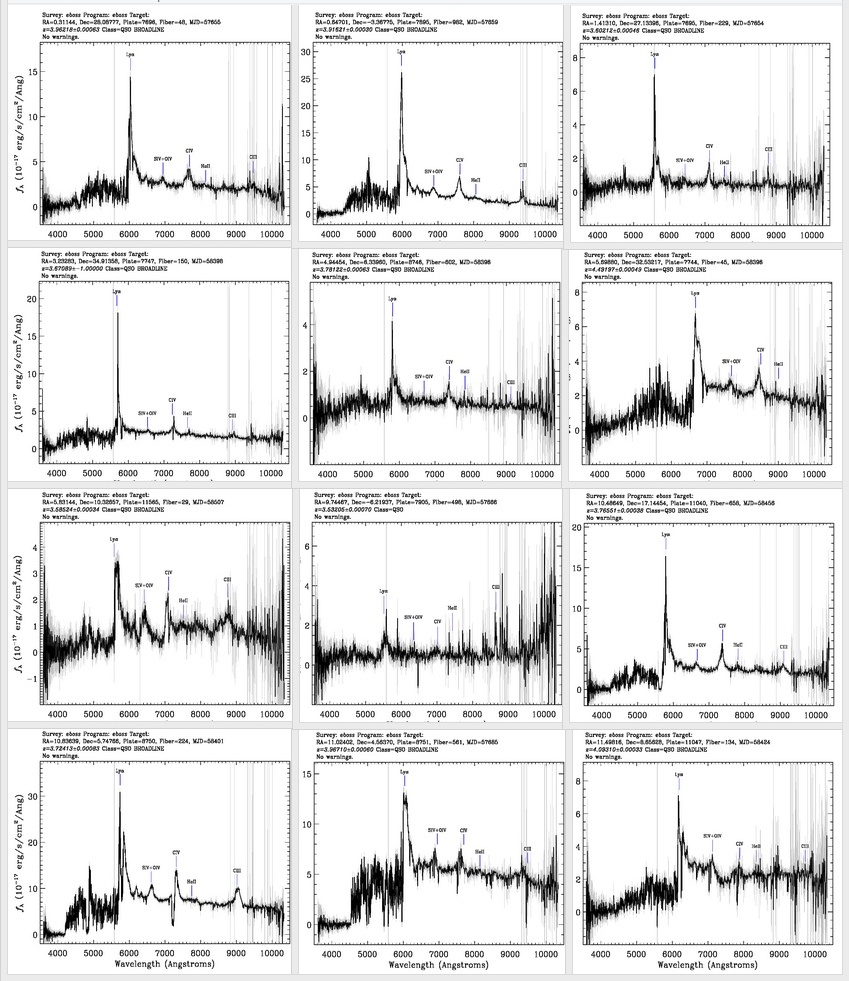} 
\caption{The first 12 of the 268 high-redshift DR16 pipeline quasars, in RA order.  Note the prominent Ly$\alpha$ line and bluewards (left) of it, all or some of the Ly$\beta$ line, the Lyman cutoff, and the Lyman forest.}
\end{figure}  

\subsubsection{\textbf{Miscellaneous}}

While the selected DR16 pipeline objects are those with populated SUBCLASS and not having ZWARNING=4, others are provisionally processed alongside.  Those provisional objects which are subsequently found to have radio/X-ray/AllWISE associations are classified as quasars in Milliquas if those associations calculate out to a pQSO of 97\%+.  The basis for this, and details of the processing, are explained in the HMQ paper \citep{HMQ}, Section 8.

Figure 12 shows 2 DR16 pipeline quasars which were identified to the wrong member of a star-quasar doublet.  Milliquas presents the correct optical objects for these quasars.

\begin{figure} 
\includegraphics[scale=0.225, angle=0]{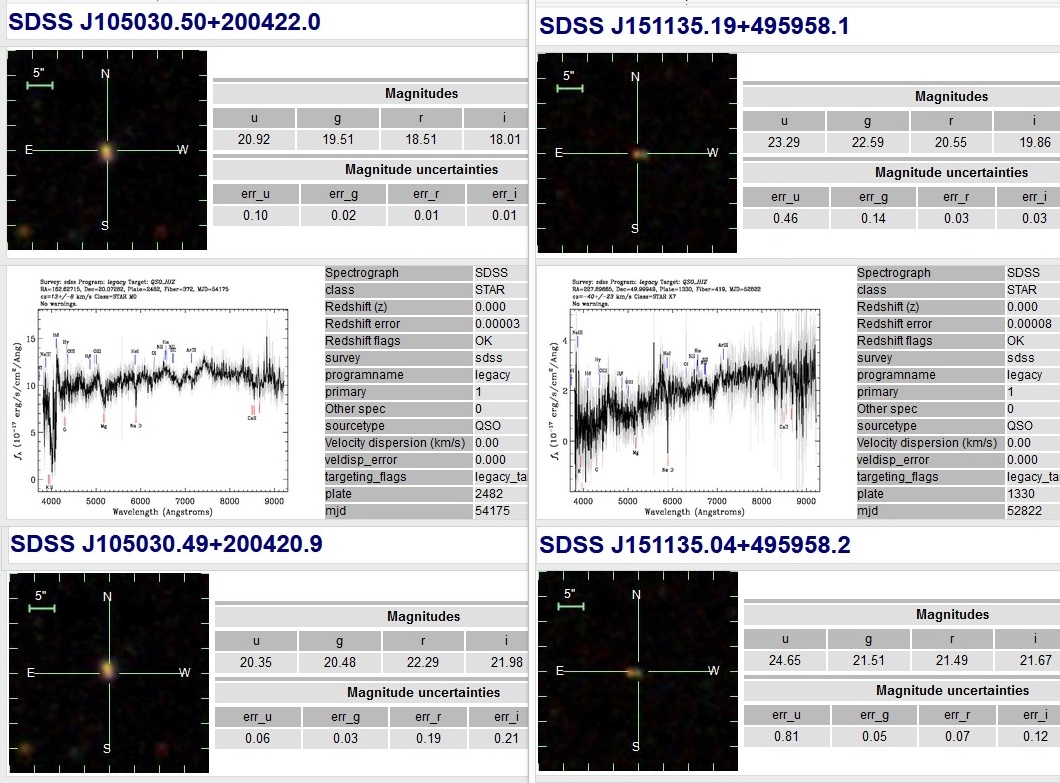} 
\caption{Two new quasars identified in close doublets with stars; DR16Q visually classified the stars as quasars due to the merged spectra.  Left half of figure shows one doublet, right half shows the other, zoom in to see clearly.  Left doublet: star cross-haired in top image shows stellar photometry, quasar is cross-haired in bottom image.  The spectrum shows quasar lines at the blue end where the quasar's flux matches the star's flux; DR16Q-given QuasarNET redshift is 1.650.  Right doublet: red star in top image shows typical red-star photometry, blue quasar in bottom image shows flat photometry with dropout in \textit{u}.  Spectrum shows a Ly$\alpha$ line at 4550\AA; DR16Q-given QuasarNET redshift is 2.709.}
\end{figure}

\begin{figure} 
\includegraphics[scale=0.225, angle=0]{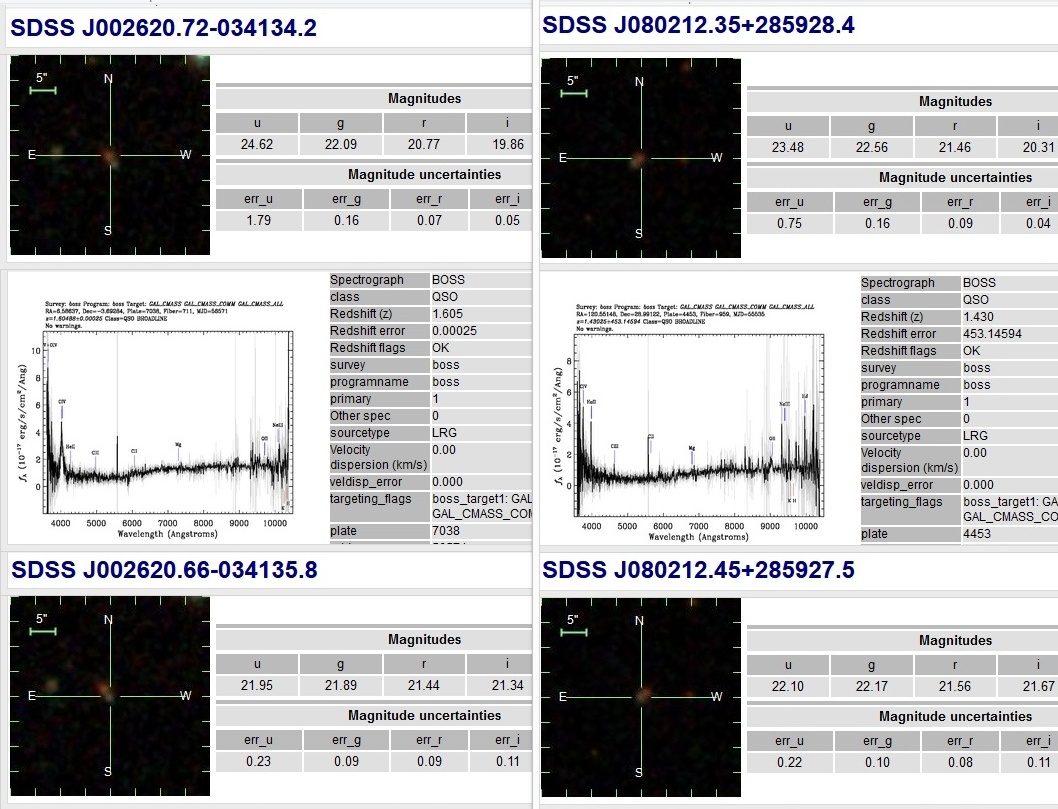} 
\caption{Two new quasars identified in close doublets with stars; left half of figure shows one doublet, right half shows the other, zoom in to see clearly.  The DR16 pipeline targeted the stars and classified them as quasars due to the merged spectra.   In both cases, the star cross-haired in top image shows \textit{ugriz} photometry typical of a red star, and the quasar cross-haired in bottom image shows flat \textit{ugriz} photometry.  The merged spectra show quasar lines at the blue end where the quasar's flux dominates the star's flux.  The pipeline redshift of the left-side object is 1.605, the right-side object has z=1.430.}
\end{figure}

\section{New Quasars}  

\citet{FLES40}, Sec. 3.1, identified 3 close star-quasar doublets in which DR16Q had identified the wrong object as the quasar; these were swapped out to identify 3 new quasars.  Here, Fig. 11 shows 2 similar star-quasar doublets in which the DR16Q-identified quasars were onto the wrong objects, which are now swapped out to select the true quasar.  These are thus 2 new quasar discoveries.  The figure caption gives more details.  

Fig. 12 is much the same, but instead of DR16Q quasars, they are DR16 pipeline quasars which are swapped out.  These are once again 2 new quasar discoveries, and the figure caption gives more details.

\section{New Quasar Candidates}  

11 radio/X-ray associated quasar candidates were found to be in close doublets with SDSS-classified stars.  The appearance is that SDSS took a spectrum there because of the radio/X-ray detection, but pointed it at the brighter star and missed the true source.  Thus the true sources have had no spectrum.  They are: 
\begin{itemize}
\item J015304.49-002144.1, z$_{phot}$=3.1, FIRSTJ015304.4-002143     
\item J020307.48-043216.6, z$_{phot}$=2.2, 4XMM J020307.5-043216    
\item J084733.18+065212.7, z$_{phot}$=2.8, FIRSTJ084733.1+065212  
\item J120518.51+210358.6, z$_{phot}$=0.6, FIRSTJ120518.5+210359   
\item J121038.70+495050.7, z$_{phot}$=1.9, 4XMM J121038.7+495051  
\item J122245.56+402126.7, z$_{phot}$=1.2, FIRSTJ122245.5+402126   
\item J132123.56+040909.3, z$_{phot}$=1.5, 4XMM J132123.5+040910   
\item J144949.26+090152.4, z$_{phot}$=1.6, CXOG J144949.2+090152  
\item J145736.04+093029.3, z$_{phot}$=0.8, FIRSTJ145736.1+093028   
\item J205101.30-004955.8, z$_{phot}$=1.4, 4XMM J205101.2-004955   
\item J222647.28+254839.5, z$_{phot}$=3.1, CXOG J222647.2+254839   
\end{itemize}

\section{Inclusion of VLASS}

The VLASS Quick Look catalogue\footnote{at https://cirada.ca/catalogues} is their first complete publication of VLASS radio sources.  The VLASS prefix of ``VLASS1QLCIR'' is abbreviated to "VL0" in Milliquas to fit into the standard Milliquas catalogue format.  The VLASS data adds 34\,189 new radio core associations and 6793 probable double radio lobe associations to the Milliquas data.  The VLASS "Quick Look" astrometry is said to be offset by up to 2 arcsec in places within $\delta\leq-10^{\circ}$ and $\delta\geq+60^{\circ}$; the Milliquas processing should accomodate most of that, as it searches for core associations within 2.5 arcsec.

\section{Conclusion} 

The Milliquas (Million Quasars) catalogue v7.2 is presented as a complete record of published quasars to 30 April 2021, including the SDSS-DR16Q quasar release.  Milliquas presents 829\,666 type 1 QSOs \& AGN, 703\,348 photometric quasars of pQSO$\geq{60\%}$, 2704 BL Lac objects, and 32\,726 type 2 objects.  Astrometry is 0.01 arcsecond accurate for most objects, and red-blue photometry is of 0.01 magnitude precision.  X-ray and radio associations for these objects are presented as applicable, including double radio lobes.  A detailed presentation is made of the SDSS-DR16Q uptake.  VLASS associations are included for the first time.

\section*{Acknowledgements}
Thanks to supportive correspondents.  Thanks to Heinz Andernach for getting me focused on VLASS.  This work was not funded.





\label{lastpage}
\end{document}